# Deep Fuzzy Systems

Khaled A. Nagaty
Faculty of Informatics & Computer Science
The British University in Egypt, ElSherouk City, Cairo, Egypt
+02-01001012413
khaled.nagaty@bue.edu.eg

**ABSTRACT-** An investigation of deep fuzzy systems is presented in this paper. A deep fuzzy system is represented by recursive fuzzy systems from an input terminal to output terminal. Recursive fuzzy systems are sequences of fuzzy grade memberships obtained using fuzzy transmition functions and recursive calls to fuzzy systems. A recursive fuzzy system which calls a fuzzy system $n$ times includes fuzzy chains to evaluate the final grade membership of this recursive system. A connection matrix which includes recursive calls are used to represent recursive fuzzy systems.

—————————◆—————————

## 1. INTRODUCTION

A fuzzy system can be seen as undirected finite graph the grade membership of the edges is in the interval [0,1]. In fuzzy systems information is transferred from input terminal A to output terminl B through a network of fuzzy chains which is represented using fuzzy transfer function (FTF). A deep fuzzy system recursively calls itself or another fuzzy system. It is in this sense that recursion is typically handled using stack. Intuitively, a recursive fuzzy system can be viewed as a procedure or module where the execution starts with the main module and the processes continue by calling fuzzy systems recursively. The recursive fuzzy system consists of single module which is the main module. The module either follows non-recursive chains from input terminal to output terminal where the intersection of the edge grade-memberships is performed to obtain the grade-membership of the chain or recursively calls its main module or another module. After returning from the call it finds the edge grade-membership of the deeper chain until it reaches the outer chain and reaches the output terminal. To choose a fuzzy chain from input terminal to output terminal it is required to return from all calls of recursive chains and reach the output terminal of the main module.

## 2. PRELIMINARIES

**Definition 2.1.**
A fuzzy system with $n$-input terminals $x_1, \ldots, x_n$ and a single output terminal $\xi$, where the fuzzy function $f$ is represented by [1]:
$$f(x_1, \ldots, x_n) = \xi \qquad (1)$$

**Definition 2.2.**
Let $x_1, x_2, \ldots, x_n$ be $n$ points in the fuzzy set $X$ with $\mu(x_i, x_j)$ be the grade membership that describes the transition from $x_i$ to $x_j$, $1 \le i, j \le n$. A sequence $f = (x_r, \ldots, x_t)$ will be said a fuzzy chain of degree one from $x_r$ to $x_t$.

## 3. Deep fuzzy chains

Let $f_1, f_2, \ldots, f_k$ be $k$ fuzzy chains of degree one in the set of degree one fuzzy chains $k$ with $\mu'(f_i, f_j)$ being the grade membership for the transition from $f_i$ to $f_j$ such that $1 \leq i, j \leq k$.

A sequence $f' = (f_r, \ldots, f_t)$ is said to be a fuzzy chain of degree two from $f_r$ to $f_t$.

Let $f_1', f_2', \ldots, f_k'$ be $k$ fuzzy chains of degree two in the set of degree two fuzzy chains $M$ with $\mu''(f_i', f_j')$ being the grade membership of the transition from $c$ to $f_j'$ such that $1 \leq i, j \leq k$.

A sequence $f'' = (f_r', f_{r+1}', \ldots, f_t')$ is considered a fuzzy chain of degree three from $f_r'$ to $f_t'$ such that $1 \leq r, t \leq k$.

In general, a deep fuzzy chain is defined as follows:

**Definition 3.1**

Let $f_1^{n-1}, f_2^{n-1}, \ldots, f_k^{n-1}$ be $k$ fuzzy chains of degree $n-1$ in the set of degree $n-1$ fuzzy chains $k$ with $\mu^n(f_i, f_j)$ being the grade membership for the transition from $f_i^{n-1}$ to $f_j^{n-1}$ such that $1 \leq i, j \leq k$.

A sequence $f^n = (f_r^{n-1}, \ldots, f_t^{n-1})$ will be said a fuzzy chain of degree $n$ from $f_r^{n-1}$ to $f_t^{n-1}$.

Fig.1 shows five fuzzy systems namely $\psi_1, \psi_2, \psi_3, \psi_4$ and $\psi_5$ with five fuzzy transmission systems $F_{\psi_1}, F_{\psi_2}, F_{\psi_3}, F_{\psi_4}$ and $F_{\psi_5}$ respectively. Each fuzzy system has four fuzzy chains of fuzzy grade memberships from the input terminal A to output terminal B.

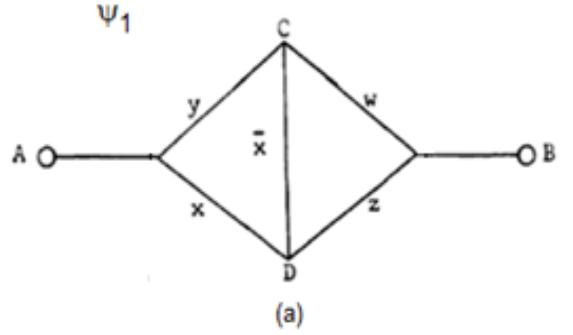
(a)

$$F_{\psi_1} = xz + x\bar{x}w + yw + y\bar{x}z \qquad (2)$$

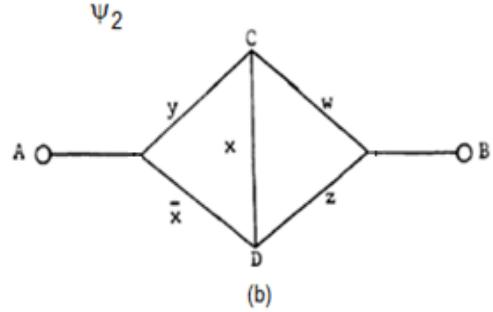
(b)

$$F_{\psi_2} = \bar{x}z + \bar{x}xw + yw + yxz \qquad (3)$$

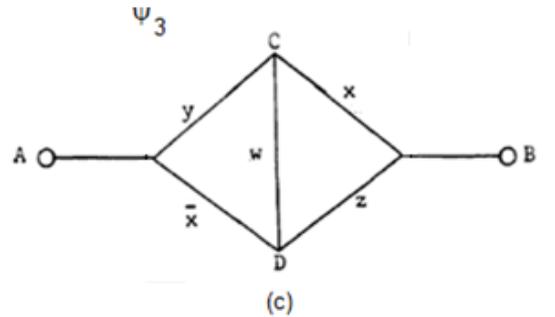
(c)

$$F_{\psi_3} = \bar{x}z + \bar{x}wx + yx + ywz \qquad (4)$$

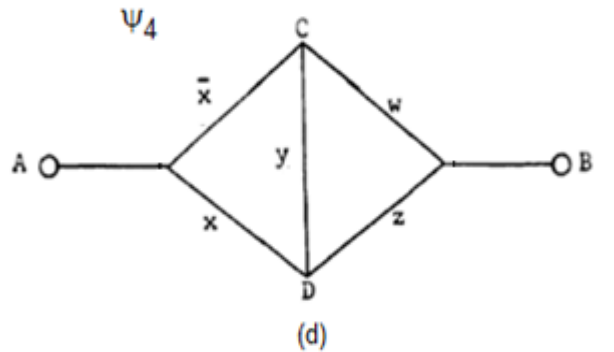
(d)

$$F_{\psi_4} = \bar{x}w + \bar{x}yz + xz + xyw \qquad (5)$$

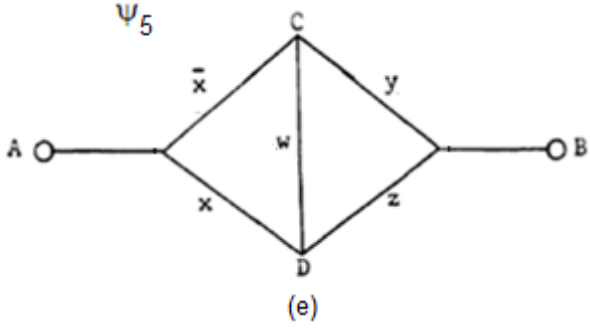

$F_{\psi_5} = \bar{x}y + \bar{x}wz + xz + xwy$ (6)

Fig.1 Five graph fuzzy systems with fuzzy transmission functions

Each entry $x_{ij}$ in the connection matrix $\rho$ for a non-recursive system is a fuzzy transmission function from entry $i$ to entry $j$. Fig.2 shows the connection matrix for the fuzzy system $\psi_1$.

|   | A | B | C | D |
|---|---|---|---|---|
| A | 1 | 0 | y | x |
| B | 0 | 1 | w | z |
| C | y | w | 1 | $\bar{x}$ |
| D | x | z | $\bar{x}$ | 1 |

Fig. 2 The connection matrix for $\psi_1$

Consider the deep two terminal fuzzy system $\Phi$ in Fig. 3 where each edge is a recursive call of fuzzy systems $\psi_1, \psi_2, \psi_3, \psi_4$ or $\psi_5$ where $\psi_1$ is recursively called $k_1$ times, $\psi_2$ is recursively called $k_2$ times, $\psi_3$ is recursively called $k_3$ times, $\psi_4$ is recursively called $k_4$ times and finally $\psi_5$ is recursively called $k_5$ times.

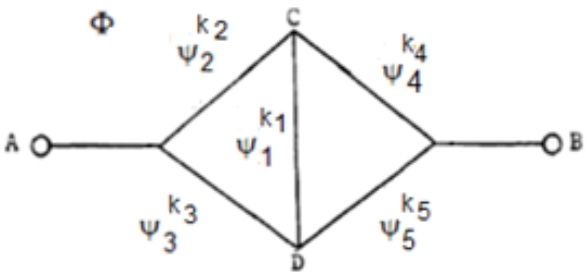

Fig. 3 Graph for a deep fuzzy system

**Definition 3.2**
Let $\Phi$ be a two-terminal fuzzy system that is constructed from edge types $\psi_1^{k_1}, \psi_2^{k_2}$, $\psi_3^{k_3}, \psi_4^{k_4}$ and $\psi_5^{k_5}$ which are the number of calls to the fuzzy systems $\psi_1, \psi_2, \psi_3, \psi_4, \psi_5$ such that $\psi_1$ is called $k_1$ times, $\psi_2$ is called $k_2$ times, $\psi_3$ is called $k_3$ times, $\psi_4$ is called $k_4$ times and $\psi_5$ is called $k_5$ times.

Fig. 4 shows the connection matrix of Fig. 3

|   | A | B | C | D |
|---|---|---|---|---|
| A | 1 | 0 | $\psi_2^{k_2}$ | $\psi_5^{k_5}$ |
| B | 0 | 1 | $\psi_4^{k_4}$ | $\psi_5^{k_5}$ |
| C | $\psi_2^{k_2}$ | $\psi_4^{k_4}$ | 1 | $\psi_1^{k_1}$ |
| D | $\psi_3^{k_3}$ | $\psi_5^{k_5}$ | $\psi_1^{k_1}$ | 1 |

Fig. 4 The connection matrix for Fig.3

The connection matrix $\rho$ for Fig. 3 is a square matrix where each entry $x_{ij}$ is recursive calls to fuzzy systems $\psi_1, \psi_2, \psi_3, \psi_4$ and $\psi_5$. A recursive fuzzy transmission is symmetric as the graph is undirected such that $x_{ij} = x_{ji}$ for all $i \neq j$ and $x_{ii} = 1$ for all $i$. In recursive call the fuzzy transmission matrix $\rho$ for fuzzy system $\psi_1$ is of order $n$. It is clear that $\rho^2$ is:

$$\sum_{t=1}^{n} \psi_{it}^k \psi_{tj}^k \quad (7)$$

This is the grade member ship of:

$$\max_t [\min(\min(\psi_{it}^k), \min(\psi_{tj}^k))] \quad (8)$$

Iff there is a direct path from vertex $i$ to vertex $j$ through $k$ recursive calls of fuzzy system $\psi$.

Example (1)
Suppose the deep two terminal fuzzy system in Fig. 5 is:

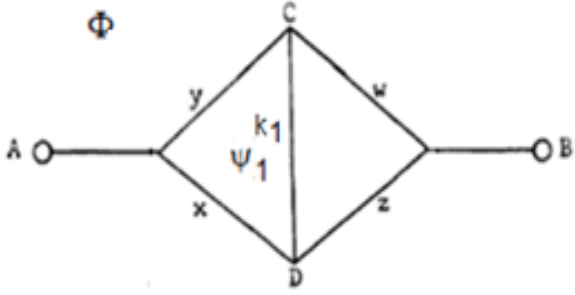

Fig.5 Graph of a two terminal fuzzy system

Fig. 6 shows the connection matrix for Fig.5 is:

|   | A | B | C | D |
|---|---|---|---|---|
| A | 1 | 0 | y | x |
| B | 0 | 1 | w | z |
| C | y | w | 1 | $\psi_1^{k_1}$ |
| D | x | z | $\psi_1^{k_1}$ | 1 |

Fig. 6 The connection matrix for Fig. 5

Assume $\psi_1$ is called $k_1$ times then $\psi_1^{k_1}$ appears in the fuzzy chains $x\psi_1^{k_1}w$ and $y\psi_1^{k_1}z$ which means that the fuzzy system $\psi_1$ is called each time it appears in the fuzzy chain. The recursive fuzzy chain system which consists of single module Φ is called the main module. The module either uses yw or xz and outputs the result to node B or calls $\psi_1$ in the fuzzy chain $y\psi_1^{k_1}z$ or $x\psi_1^{k_1}w$ and after returning from all calls it outputs the result to terminal B. To submit the output to terminal B we require that all output must return from all calls. The fuzzy transmission function FTF of Φ is defined as the concatenations of the union of the closed chains between the terminals of the system. The FTF of the deep two-terminal fuzzy system Φ of Fig. 2 is:

$$F_\Phi = \psi_2^{k_2} \psi_4^{k_4} + \psi_2^{k_2} \psi_1^{k_1} \psi_5^{k_5} + \psi_3^{k_3} \psi_1^{k_1} \psi_4^{k_4} + \psi_3^{k_3} \psi_5^{k_5} \quad (9)$$

Where:

+ represents max and concatenation represents min operations.

**Algorithm 1**
1) Determine all redundant input-output deep fuzzy chains.
2) For each fuzzy system call $\psi_i^{k_i}$ in a deep fuzzy chain do:
   2.1 Call the fuzzy system $\psi_i$ a $k_i$ times using algorithm 2.
   2.2 If $k_i \leq 0$ and NOT end of fuzzy systems calls get the next call $\psi_i^{k_i}$ in the current fuzzy chain and go to step i.
   End For
3) If NOT end of deep fuzzy chains then get the next deep fuzzy chain and go to step 2).
4) Form the union of all deep fuzzy chains.
5) Find the max of all grade-memberships of the chains obtained in iii)
6) Input the output grade-membership from iv) to the next input of the deep fuzzy chain.
7) Repeat step 1).

**Example (2)**

Suppose in eq.(9) of the fuzzy system Φ we have the deep fuzzy chain $y\psi_1^2 z$ of the fuzzy system Φ in Fig.1. This deep chain calls the fuzzy system $\psi_1$ two times. In the first call in $y\psi_1 z$ the returning point z is pushed on top of the stack. Suppose in the second call for $\psi_1$ the fuzzy chain $x\psi_1 w$ is obtained then w is pushed on top of the stack before $\psi_1$ is called. Suppose the output of $\psi_1$ of the second call is xz or yw which means that there is no other call for $\psi_1$. At this point the returning point on top of the stack i.e. w is poped and the ouput of $x\psi w$ is calculated which is xxzw + xyww. The stack is popped again and $y\psi_1^2 z$ is calculated which is:

y(xxzw + xyww)z.

## Algorithm 2

Without loss of generality considers the two terminal deep fuzzy system $\psi_1^{k_1}$ in Fig. 4 [1] where the fuzzy system $\psi_1$ is recurrently called $k_1$ times, where the output of terminal B becomes the input to terminal A.

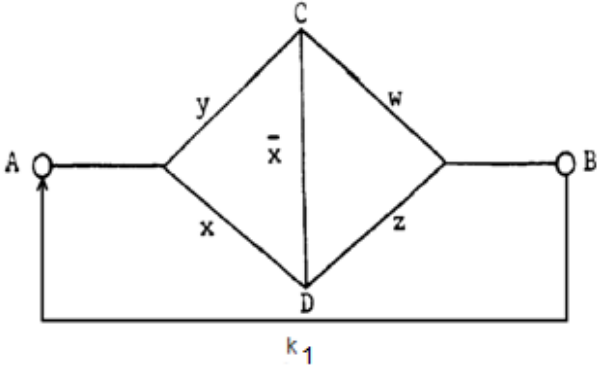

Fig. 4 Graph of recurrent fuzzy system $\psi_1^{k_1}$

$$F_{\psi_1}^{k_1} = (xz + x\bar{x}w + yw + y\bar{x}z)^{k_1} \quad (10)$$

Let:

$x_1 = xz,\ x_2 = x\bar{x}w,\ x_3 = yw,\ x_4 = y\bar{x}z$

Where + represents max and concatenation represents min operations.

$$F_{\psi_1}^{k_1} = \sum_{n_1+n_2+n_3+n_4=k_1} \binom{k_1}{n_1,n_2,n_3,n_4} \prod_{i=1}^{4} x_i^{n_i} \quad (11)$$

Where:

$$\binom{k_1}{n_1,n_2,n_3,n_4} = \frac{k_1!}{n_1!n_2!n_3!n_4!} \quad (12)$$

Where $\Sigma$ represents max and $\Pi$ represents min operations.

Therefore, the FTF of the deep two-terminal fuzzy system calls is obtained as follows:

For each fuzzy system call $\psi_i^{k_i}$ do
1) Determine all redundant input-output fuzzy chains.
2) For each fuzzy chain do
   2.1 Form the concatenation of the corresponding edge grade-memberships of the chain.
   2.2 Find the min of the obtained concatenation in i) to obtain the grade-membership of the chain.
   2.3 Form the union of all obtained chains.
   2.4 Find the max of all grade-memberships of the chains obtained in iii)
3) Input the output grade-membership from 2.4) to the next input of the deep fuzzy chain.
4) Repeat step 1).

## Algorithm 3

This algorithm is based on Warshall algorithm in [2] to compute the transitive closure of a recursive binary matrix.
1. Label all vertices of the two-terminal fuzzy system $\psi$ by the integers $1, \ldots, N$.
2. Construct the recursive connection matrix $\rho_\psi$ for the two-terminal fuzzy system $\psi$ connecting vertex $i$ to vertex $j$ through a direct chain or recursive chain.

```
FOR K = 1 TO N
FOR I = 1 TO N
FOR J = 1 TO N
IF (ρ_ψ(I,K) ≠ ψ_T^R) && (ρ_ψ(I,K) ≠ 0)
THEN
  ρ_ψ(I,J)
  = max (ρ_ψ(I,J), min (ρ_ψ(I,K), ρ_ψ(K,J)))
END
IF (ρ_ψ(I,K) = ψ_T^R ) THEN
 FOR T= R TO 1
ρ_ψ(I,K)
= RETURN (max(ρ_ψ(I,K), min( ρ_ψ(I,1), ρ_ψ(1,K)))
END
END
END
END
```

The basic idea is to process all the loops within column $K$ such that:

$$\rho_\psi(I,J) = M(I,J,TK) \quad (13)$$

Where:
$M(I,J,TK) \triangleq$
$max\{min(\text{all chains from } I \text{ to } J$
$(min(\text{all chains with loops from } I \text{ to } J))\}$
such that each intermediate node has label ≤ K and number of loops ≤ T.

At the termination of the algorithm we note that

$$\rho_\psi(I,J) = M(I,J,NR) \quad (14)$$

Where:
$M(I,J,K) \triangleq$
$max\{min(\text{all chains from } I \text{ to } J$
$(min(\text{all chains from } I \text{ to } J))\} \quad (15)$

This assertion can be proved using induction as shown in [1]:

1) K=1, $T = 1$ there is no other call to a fuzzy system which makes the depth equals to 1 and the original matrix will be:
   $M(I,J,1)$
   $= (max(\rho_\psi(I,J), min(\rho_\psi(I,1), \rho_\psi(1,J)))$
2) Assume $\rho_\psi(I,J) = M(I,J,TK)$ which means that a fuzzy system is called $T$ times. It is required to show that:
   $\rho_\psi(I,J) = M(I,J,TK+1))$
3) If $M(I,J,TK+1))$ does not involve $TK+1$ then the matrix is $\rho_\psi(I,J) = M(I,J,TK)$ as the assumption in step 2.
4) If $M(I,J,TK+1))$ does involve $TK+1$, then:
   $M(I,J,TK+1)) = M(I,J,TK) + M(I,J,1)$

Since $M(I,J,1)$ is true from step 1 and $M(I,J,TK)$ is true from the assumption of step 2 then we conclude that $M(I,J,TK+1))$ is true.

### 4. CONCLUSION

Deep fuzzy systems have been introduced and a conceptual framework for studying deep fuzzy systems is provided.

Type equation here.